\documentclass[sigconf]{acmart}

\usepackage{stfloats}

\AtBeginEnvironment{quote}{\itshape}
\AtBeginDocument{%
\providecommand\BibTeX{{%
\normalfont B\kern-0.5em{\scshape i\kern-0.25em b}\kern-0.8em\TeX}}}

\newcommand{\rr}[1]{\textcolor{black}{#1}}

\copyrightyear{2024}
\acmYear{2024}
\setcopyright{rightsretained}
\acmConference[DIS '24]{Designing Interactive Systems Conference}{July 1--5, 2024}{IT University of Copenhagen, Denmark}
\acmBooktitle{Designing Interactive Systems Conference (DIS '24), July 1--5, 2024, IT University of Copenhagen, Denmark}\acmDOI{10.1145/3643834.3660710}
\acmISBN{979-8-4007-0583-0/24/07}

\acmSubmissionID{5597}

\begin{document}

\title[Designing a conversational telepresence robot for homebound older adults]{``This really lets us see the entire world:'' Designing a conversational telepresence robot for homebound older adults}





\author{Yaxin Hu}
\orcid{0000-0003-4462-0140}
\affiliation{%
  \institution{\normalsize{Department of Computer Sciences\\University of Wisconsin--Madison}}
  \streetaddress{\normalsize{University of Wisconsin--Madison, Madison, Wisconsin}}
  \country{} 
}
\email{yaxin.hu@wisc.edu}

\author{Laura Stegner}
\orcid{0000-0003-4496-0727}
\affiliation{%
  \institution{\normalsize{Department of Computer Sciences\\University of Wisconsin--Madison}}
  \streetaddress{\normalsize{University of Wisconsin--Madison, Madison, Wisconsin}}
  \country{} 
  }
\email{stegner@wisc.edu}

\author{Yasmine Kotturi}
\orcid{0000-0001-6201-7922}
\affiliation{ 
  \institution{\normalsize{Human-Computer Interaction Institute\\Carnegie Mellon University}}
  \streetaddress{\normalsize{Carnegie Mellon University, Pittsburgh, Pennsylvania}}
  \country{} 
  }
\email{ykotturi@andrew.cmu.edu}

\author{Caroline Zhang}
\orcid{0009-0000-8737-5929}
\affiliation{%
  \institution{\normalsize{Department of Mathematics\\University of Wisconsin--Madison}}
  \streetaddress{University of Wisconsin--Madison, Madison, Wisconsin}
  \country{} 
  }
\email{mzhang477@wisc.edu}

\author{Yi-Hao Peng}
\orcid{0000-0002-6335-5904}
\affiliation{ 
  \institution{\normalsize{Human-Computer Interaction Institute\\Carnegie Mellon University}}
  \streetaddress{Carnegie Mellon University, Pittsburgh, Pennsylvania}
  \country{} 
  }
\email{yihaop@cs.cmu.edu}

\author{Faria Huq}
\orcid{0000-0002-6247-8283}
\affiliation{ 
  \institution{\normalsize{Human-Computer Interaction Institute\\Carnegie Mellon University}}
  \streetaddress{Carnegie Mellon University, Pittsburgh, Pennsylvania}
  \country{} 
  }
\email{fhuq@andrew.cmu.edu}

\author{Yuhang Zhao}
\orcid{0000-0003-3686-695X}
\affiliation{ 
  \institution{\normalsize{Department of Computer Sciences\\University of Wisconsin--Madison}}
  \streetaddress{\normalsize{University of Wisconsin--Madison, Madison, Wisconsin}}
  \country{} 
  }
\email{yuhang.zhao@cs.wisc.edu}

\author{Jeffrey P. Bigham}
\orcid{0000-0002-2072-0625}
\affiliation{ 
  \institution{\normalsize{Human-Computer Interaction Institute\\Carnegie Mellon University}}
  \streetaddress{Carnegie Mellon University, Pittsburgh, Pennsylvania}
  \country{} 
  }
\email{jbigham@cs.cmu.edu}

\author{Bilge Mutlu}
\orcid{0000-0002-9456-1495}
\affiliation{ 
  \institution{\normalsize{Department of Computer Sciences\\University of Wisconsin--Madison}}
  \streetaddress{University of Wisconsin--Madison, Madison, Wisconsin}
  \country{} 
  }
\email{bilge@cs.wisc.edu}

\renewcommand{\shortauthors}{Hu, et al.}

\begin{abstract}

In this paper, we explore the design and use of \textit{conversational telepresence robots} to help homebound older adults interact with the external world. An initial needfinding study (N=8) using video vignettes revealed older adults' experiential needs for robot-mediated remote experiences such as exploration, reminiscence and social participation. We then designed a prototype system to support these goals and conducted a technology probe study (N=11) to garner a deeper understanding of user preferences for remote experiences. \rr{The study revealed user interactive patterns in each desired experience, highlighting the need of robot guidance, social engagements with the robot and the remote bystanders. Our work identifies a novel design space where conversational telepresence robots can be used to foster meaningful interactions in the remote physical environment. We offer design insights into the robot's proactive role in providing guidance and using dialogue to create personalized, contextualized and meaningful experiences.}
\end{abstract}


\begin{teaserfigure}
  \includegraphics[width=\textwidth]{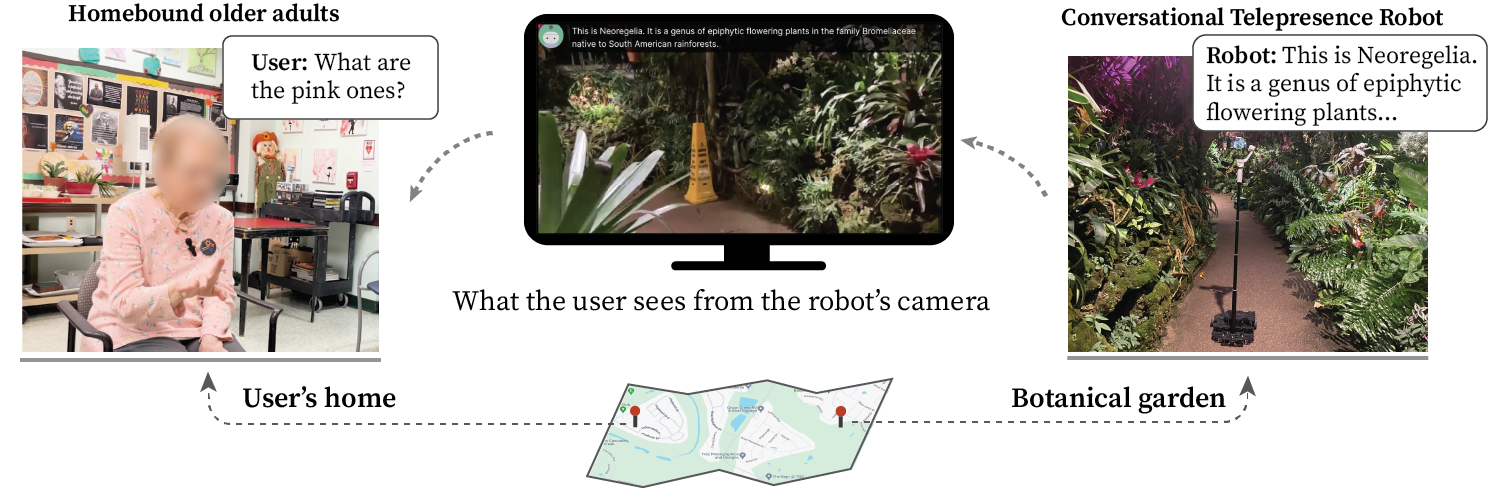}
  \vspace{-15pt}
  \caption{In this paper, we explore how \textit{conversational telepresence robots} might address the needs and expectations of homebound older adults in experiencing the world outside their homes. Through a \textit{needfinding study}, participants articulated two experiences of interest for robotic telepresence: exploration and reminiscence. Then, we prototyped a telepresence robot to support these experiences and conducted a \textit{technology probe study} to understand users' preferences with robotic telepresence.}
  \vspace{9pt}
  \Description{Overview of an interaction scenario. The image portrays a communication exchange between a homebound older adult staying at home and a telepresence robot situated in a botanical garden. The older adult inquires the robot about the name of the flora that is observed through the robot's camera (left). The central part of the image provides a visual example of what the user sees on their screen as captured by the robot's camera (center). Receiving the question from older adults, the robot identifies and shares information about the plants to the older adult (right).}
  \label{fig:teaser}
\end{teaserfigure}

\maketitle

\section{Introduction}


The past 30 years has seen the emergence of telepresence robots designed to connect people with places that they could not reach and engage in social interactions with people at a distance \cite{minsky1980telepresence, paulos1996delivering, paulos1998prop}. These technologies are often equipped with a mobile base and can be controlled by the remote user to navigate in the environment in which the robot is placed. Prior research has identified great potential in using telepresence robots for older adults at home \cite{tsai2007developing, bevilacqua2014telepresence, beer_mobile_2011} and in care facilities \cite{koceska_telemedicine_2019, niemela2021towards, arthanat2023perspectives, broadbent2016benefits}. Example use cases of telepresence robots for older adults include remote medical appointments \cite{koceska_telemedicine_2019, bakas2018satisfaction, broadbent2016benefits}; communication with family members and caretakers \cite{cesta_long-term_2016, koceski_evaluation_2016, moyle_connecting_2014, moyle2017potential, fiorini2022living}; task management \cite{coradeschi2013giraffplus}; remote education \cite{hiyama2017facilitating}; and health monitoring \cite{cortellessa2018robin, coradeschi2013giraffplus}. Telepresence robots can be particularly helpful for \textit{homebound} older adults to access places they want to go in the external world even when they are constrained in their dwelling environment. Compared to the general elderly population, homebound older adults have a significantly higher risk of mortality \cite{cohen2010Effect} and suffer more from functional impairments and mental illnesses \cite{Qiu2010Physical, Ornstein2015Epidemiology, choi2007comparison, Soones2017TwoYear}. In the last decade, the prevalence of homebound adults who are aged 70 years or older has more than doubled, increasing from  5.0\% to 13.0\%, and this number continues to increase \cite{ankuda2021association}. 




\rr{The majority of the existing research used telepresence robots for functional tasks and communication purposes, and very little is known about how telepresence robots might enable older adults to explore and experience remote physical environments. Furthermore, while existing research advocated that the robot should take an invisible role in mediating the communications \cite{tsui_accessible_2015}, we aid to study the robot's proactive role in facilitating the remote experiences through its dialogue capability.} Specifically, we explore the following research questions: (\textbf{RQ.1}) \textit{what are homebound older adults' expectations with respect to experiencing the external world}; (\textbf{RQ.2}) \textit{how might telepresence robots be designed to explore and interact with the external world}; and (\textbf{RQ.3}) \textit{what might be their experience with exploring and interacting with the external world via a telepresence robot}? \label{research questions}

\rr{To address these questions, we take a \textit{research through design (RtD)} \cite{zimmerman2007research} approach to identify design opportunities through a needfinding study, design and prototype artifacts that are informed by the findings from this study, and generate design knowledge about opportunities for future design through a technology probe study.} In the first study, we used scenario-based video vignettes as probes and conducted semi-structured interviews with eight homebound older adults to understand their experiential needs in the use of telepresence robots. From the needfinding study, we identified older adults' needs of reminiscent experience, exploratory experience and social participation through the telepresence robots. We translated these findings into design insights for a \textit{conversational telepresence robot}. Our second study involved the use of the prototype conversational telepresence robot, controlled through Wizard of Oz, as a technology probe to further understand participants' preferences for and interaction patterns within the experience. Following an on-boarding session, in two study sessions, 11 participants remotely visited a lakefront park or a botanical garden. The botanical garden is a local landmark that most participants had been to when they were younger, whereas the lakefront park is next to a university campus and none of our participants had been there before. In each remote session, participants experienced three phases: exploration with the robot's guidance, small talk with the robot, and engagement with a remote bystander. After each phase, participants were asked to reflect on their experience through semi-structured interviews. We conducted thematic analysis of the data from the interviews and the dialogue between participants and the robot prototype.

Our findings revealed that most of our participants preferred the robot's guidance in the experience over guiding the experience themselves. Participants also reminisced and disclosed personal stories when chatting with the robot and with bystanders (\textit{i.e.,} \textit{reminiscent} experience). In addition, participants viewed the robot as a guide and obtained environmental knowledge through the robot's narratives and answers to their questions (\textit{i.e.,} \textit{exploratory} experience). \rr{The findings highlighted homebound older adults' positive experiences with our novel system including the immersive, personalized and interactive experiences and the ease of access to the external world. Our participants also reported barriers in the interaction including challenges in the robot control and verbal interactions, the confusion about the robot's presence and the difficulty in comprehending the experiences. Based on our findings, we generated design implications in supporting exploratory experience, reminiscent experience and social participation for homebound older adults via the conversational telepresence robot, and highlighted the need of the robot's proactive role in guiding the the remote experiences, the use of dialogues to augment the experiences \cite{kim2024understanding}, and the social engagement to facilitate meaningful experiences. }






This work makes the following contributions:
\vspace{-2.5pt}

\begin{itemize}
    \item \textit{Design Insight:} The need for exploratory, reminiscent and social experiences in the homebound older adult population through a needfinding study;
    \item \textit{Artifact:} A conversational telepresence robot prototype for remote exploration for homebound older adults;
    \rr{\item \textit{Design Implications:} Pointing to the need of the robot's proactive role and dialogue in providing curated and personalized experiences, fostering the user's personal meaning-making, and facilitating social participation in the remote experiences;  
    \item \textit{Research through Design (RtD):} Illustration of how an \textit{RtD} approach can identify, design for, and generate knowledge from a novel space for technology design for older adults.}
\end{itemize}




\section{Related Work}

\subsection{Challenges \& Needs of Homebound Older Adults}
\textit{Homebound} means that the individual has trouble leaving home without assistive devices or help from other people because of an illness or injury as defined by the \textit{U.S. Centers for Medicare \& Medicaid Services} \cite{Medicare2018Home, US2006Glossary}. Research studies have used self-reported degrees of confinement to define homebound status and considered the participants as being homebound if they never or rarely leave the home in the last month \cite{ankuda2021association, Ankuda2021dynamics, ankuda2022experience, cohen2010Effect, Ornstein2015Epidemiology, ornstein2020estimation}, go outdoors every few days or less \cite{Sakurai2018Co, Lee2020Multidimensional}, or leave home less than once a day \cite{Fujiwara2017Synergistic}. While homebound status is often caused by injury or illness, older adults can become homebound due to environmental factors such as a pandemic, lack of transportation options, or the availability of caregivers. Older adults with homebound status have a significantly higher risk of mortality and functional decline and often suffer from multiple chronic conditions, cognitive impairments and depression at a higher rate than the non-homebound elderly population \cite{Qiu2010Physical, Sakurai2018Co, Fujita2006Frequency, Herr2013Homebound, cohen2010Effect, ornstein2020estimation, Soones2017TwoYear}. Prior research has suggested technological solutions to alleviate feelings of social isolation in homebound individuals \cite{Cohen2021Adequacy}. However, homebound older adults face more challenges in learning new technology than non-homebound older adults \cite{ankuda2022experience}. Our work aims to better understand the needs and challenges of this important and underserved population toward exploring novel technological solutions to these needs and challenges.

\subsection{Enrichment Activities for Older Adults}
Subjective wellbeing is closely associated with enrichment and leisure  activities for older adults \cite{adams_critical_2011, morse_creativity_2021} and technologies are widely studied to provide opportunities for enrichment and meaningful activities for older adults at home and in the care settings \cite{lazar_successful_2017, zhao2023older, zhao2024older, waycott2022technology}. Existing literature has broadly studied older adults' motivations and current approaches for enrichment activities \cite{waycott2022technology, lazar_successful_2017}, as well as investigated specific technologies such as virtual reality \cite{waycott2022role, baker2020evaluating} and televisions \cite{hertzog2008enrichment} for supporting these activities. \citet{waycott2022technology} surveyed a range of digital technologies used for enrichment in aged care facilities and older adults' needs to use tools such as virtual reality, Google Earth to connect to the external world and visit places they were not able to access. \rr{In particular, reminiscence is found to be closely related to older adults' subjective wellbeing through activities such as storytelling, blogging, creative expressions and oral interviews \cite{pasupathi_age_2003, westerhof2010reminiscence}. Prior research found that technology can trigger reminiscent experiences \cite{lazar2014systematic}. For instance, \citet{webber2021virtual} used digital mapping technologies for older adults to visit places of personal significance virtually and found that the reminiscence extended beyond the physical places to interpersonal relationships, cultural experiences, and even world views related to their personal past. In this work, we investigate the design and use of telepresence robots to connect older adults with the external world. Different from virtual visits through Google map or virtual reality, telepresence robots allow for access to the events and activities in the physical environment and enable social interactions with other people in real time.}

\subsection{Telepresence Robots for Older Adults and Accessibility}

Telepresence robots have been studied to increase social communication between older adults and their family members \cite{cesta_long-term_2016, mitzner_acceptance_2017, moyle_connecting_2014}, support the medical communications in the care and clinical settings \cite{hung_facilitators_2022, koceska_telemedicine_2019, koceski_evaluation_2016, clotet_assistant_2016}, and visit off-site places such as museums and sporting events \cite{beer_mobile_2011}. Telepresence robots are especially helpful for older adults isolated during the health-emergency lockdown \cite{isabet2021social}. \citet{mitzner_acceptance_2017} investigated older adults' experience of a telepresence robot for social communication and suggested design opportunities such as the robot's height, volume, size of the screen and etiquette. Factors facilitating the telepresence robot use include the increased physical presence of the remote user, free navigation in the space \cite{cesta_long-term_2016, hung_facilitators_2022}, and barriers for using the telepresence robot include privacy concerns, cost of the robot, internet connectivity \cite{beer_mobile_2011, niemela_towards_2021, hung_facilitators_2022}. Prior work also studied telepresence robots for people with cognitive or motor impairments \cite{zhang_telepresence_2022, tsui_accessible_2015, ng_cloud_2015, friedman_using_2018, tsui_designing_2013, tsui_towards_2014, barbareschi_i_2023}. These robots were used for the remote user to visit museums and galleries \cite{tsui_accessible_2015, ng_cloud_2015, friedman_using_2018}, go shopping \cite{tsui_designing_2013, tsui_towards_2014}, and work remotely for a café \cite{barbareschi_i_2023}. Telepresence robot has been reported to address the challenges of transportation for people with developmental challenges \cite{friedman_using_2018} and to increase agency for disabled teleworkers in a café \cite{barbareschi_i_2023}. \rr{In particular, \citet{tsui_designing_2013} has studied speech interfaces for telepresence robot use by people with disabilities and generated design guidelines for speech-based interfaces and highlighted the use of simple commands and design for the robot's feedback. Different from prior work's focus on user control of the robot, our work emphasizes the significance of the robot's agency and proactive role in mediating the interaction such as providing guidance, facilitating social communications, and using dialogue to support the meaning-making process during the experience.}

\subsection{Technologies for Remote Presence and Mobility}

Prior research has explored systems, methods, and mechanisms to support remote presence and mobility to connect geographically distributed users \cite{inkpen2013experiences2go}, experiences of remote locations \cite{kashiwabara2012teroos}, sense of presence of remote users in these locations \cite{adalgeirsson2010mebot, rae2013body}, as well as studied social norms associated with remote presence and mobility \cite{bhat2022confused, pfeil_bridging_2021}. These technologies have included livestreaming through mobile phone apps \cite{lu_vicariously_2019, lu2018you}, body-worn cameras \cite{neustaedter_shared_2020, neustaedter_mobile_2020, pfeil2019bodywornheigh, rae2015framework}, telepresence robots \cite{rae2017robotic, neustaedter2016beam, neustaedter2018being, uriu_generating_2021, cha2017designing}, camera glasses \cite{nicholas_friendscope_2022}, 3D mobile augmented reality systems \cite{feiner1997touring}, and drones \cite{jones_elevating_2016, shakeri_teledrone_2019}. In particular, research has explored how robotic technology can facilitate the sharing of experiences at a distance with various robot forms, such as standalone mobile robots \cite{heshmat_geocaching_2018, uriu_generating_2021, rae2015framework} and wearable robot avatars \cite{kashiwabara_teroos_2012, manabe2020exploring, kratz_polly_2014, kimber_polly_2015}. Contexts of use in which telepresence robots were explored included friends sharing leisure time outdoors \cite{heshmat_geocaching_2018}, shopping with a loved one \cite{yang2018shopping}, attending funerals \cite{uriu_generating_2021} and visiting museums and cultural heritage sites \cite{ng_cloud_2015, burgard1999experiences, thrun1999minerva}. Built on the large body of literature, we identified the gap of studying homebound older adults' exploration and experience in the physical environment using telepresence robots \rr{where the telepresence robot takes a proactive and guiding role in the interaction. To address the research questions defined in \S\ref{research questions}, we take a \textit{RtD} approach where we first conduct a needfinding study to understand homebound older adults' needs for remote experiences. Then we use the findings to inform the design of a prototype system and conduct a technology probe study to further outline this novel design space of using conversational telepresence robots to support remote experiences and social participation.}

\begin{table}[!t]
    \caption{Demographic information for our participants in the needfinding study. All participants reported having trouble leaving home without help or that leaving home is not recommended due to health conditions. }
    \label{tab:study 1 demographic}
    \centering
    \renewcommand{\arraystretch}{1.2}
    \small
    \begin{tabular}{p{0.2\linewidth}p{0.15\linewidth}p{0.15\linewidth}p{0.3\linewidth}}
    \toprule
        \textbf{ID} & \textbf{Age} & \textbf{Gender} & \textbf{Time spent living in the care facility} \\ 
        \midrule
        S1.P1 & 93 & Male & 11 years \\
        S1.P2 & 78 & Female & 4.5 years \\
        S1.P3 & 94 & Female  & 4 months \\
        S1.P4 & 91 & Female & 4 years \\
        S1.P5 & 91 & Male & 4 years \\
        S1.P6 & 91 & Female  & 4 years \\
        S1.P7 & 83 & Female  & 2.5 years \\
        S1.P8 & 84 & Female  & 8 months \\
        S1.P9 & 82 & Female  & 3 years \\
        \bottomrule
    \end{tabular}
\end{table}

\begin{figure*}[!b]
    \centering
    \includegraphics[width=\textwidth]{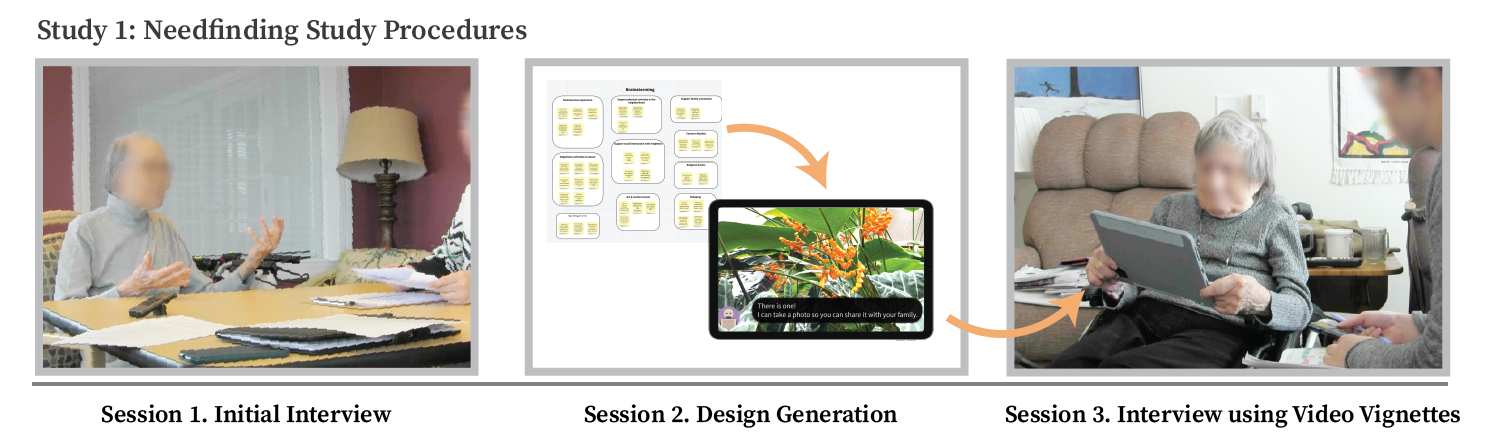}
    \caption{Overview of the needfinding study. We interviewed older adults about their homebound challenges and things they would like to do at 11 selected locations if they could go there (left). Then, we took those responses and used them to generate designs for the conversational telepresence robots (center). Finally, we presented the designs back to the participants in the form of video vignettes, then asked their feedback and discussed the use of the telepresence robots more generally (right).}
    \Description{The image is a visual summary of the needfinding study procedures. The first panel on the left shows an older adult and a researcher during an interview. The central panel illustrates the brainstorming and design phase, where ideas for remote experience scenarios are generated and represented through video vignettes. The panel on the right depicts a follow-up interview with an older adult reviewing a tablet, providing feedback on the video scenarios from step 2.}
    \label{fig:needfinding method}
\end{figure*}

\section{Study 1: Needfinding to Understand Experiential Needs of Homebound Older Adults}
We first conducted a needfinding study to answer the following research questions: \textit{(1) \textit{what are homebound older adults' expectations with respect to experiencing the external world}; (2) \textit{how might telepresence robots be designed to explore and interact with the external world}?} In particular, we aim to understand the desired experiences that older adults want to have through the robot and the desired interactions that the user wants to engage with the robot. Below, we present the needfinding study method and findings, as well as a discussion of how these findings inform Study 2.

\subsection{Method}
To address these research questions, we conducted a two-session interview study with nine participants. The insights gained in the first session were used to design video vignettes which were used as stimuli during the second session.

\subsubsection{Participants}
We recruited nine participants (seven females, two males) aged 78--94 ($M=74.27$, $SD=6.93$) from a senior living facility in the Midwestern United States (Table \ref{tab:study 1 demographic}). One participant (P4) withdrew after the first session due to medical reasons. All participants self-reported using at least one mobility aid, \textit{\textit{e.g.,}} a wheelchair or walker, and required assistance to leave the facility.  Participants were compensated \$20 USD per hour for their participation. All study sessions were video and/or audio recorded. Study materials and procedures were approved by the university's Institutional Review Board (IRB). 

\subsubsection{Study Procedure}

\paragraph{Session 1: Initial Interview}
In the first interview session, we first asked participants about their day-to-day activities and presented each participant with text prompts of 11 locations selected from OpenStreetMap taxonomy \cite{wikiopenstreetmap}: Urban Center, Restaurant, Art \& Culture, Waterfront, Park \& Garden, Market, Entertainment \& Nightlife, Outdoor Adventure, Shopping Center, Religious Location, and Sporting Events. The 11 locations are a subset of the OpenStreetMap taxonomy which are representative destinations where people spend their leisure time and therefore may be points of interest for homebound older adults to explore. Prompts were shown and discussed one by one in a consistent order for all participants. After showing the participant the prompt of each location and confirming that they had an idea of what the location was like, we asked participants what they would like to do if they were at the location as well as experiential details such as the sights, smells, sounds, and feelings they would like or dislike in each location. 

\begin{figure*}[!tb]
    \centering
    \includegraphics[width=\linewidth]{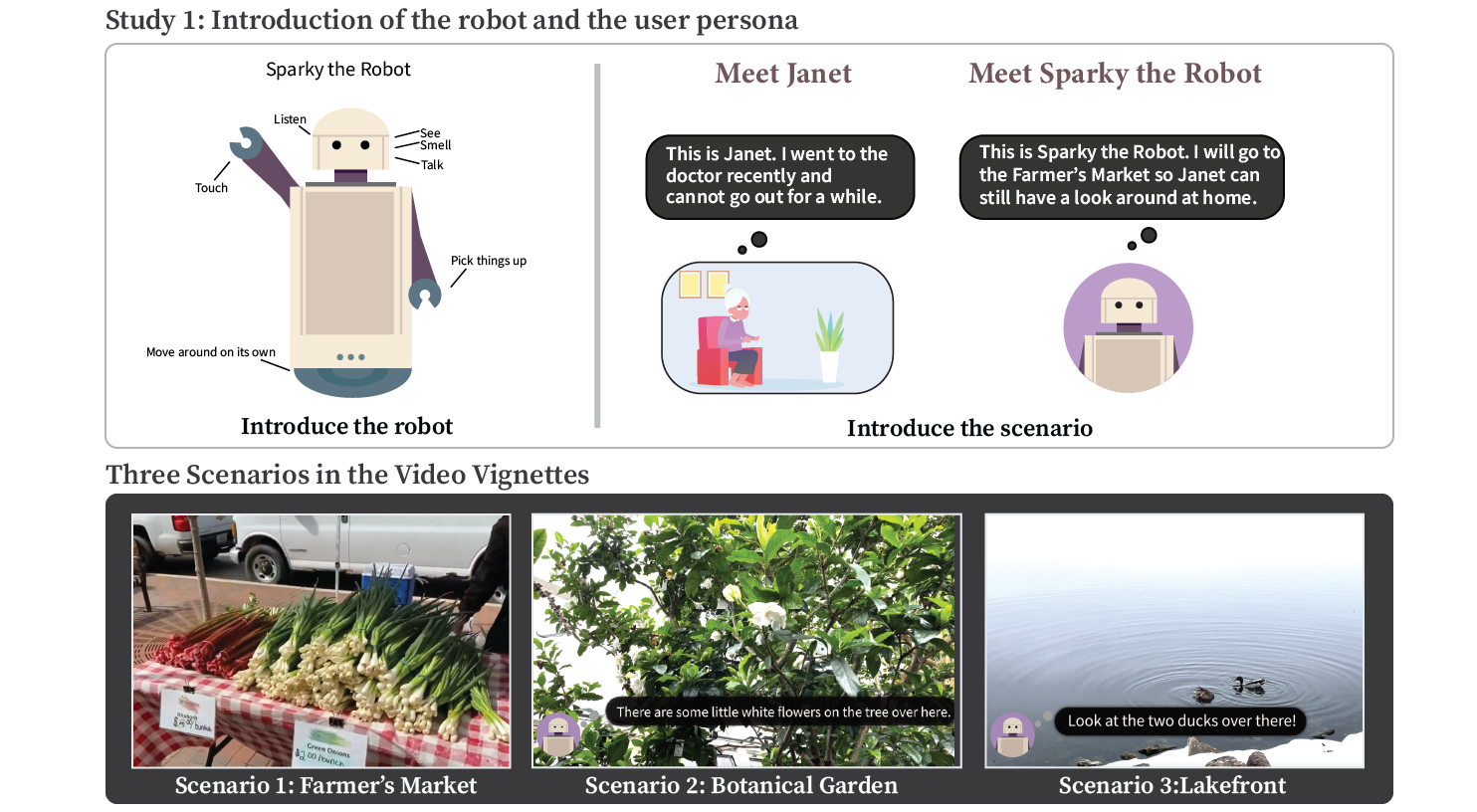}
    \caption{An overview of our needfinding study materials. We created a robot character used in our video vignettes. The robot has advanced sensing, navigation, and manipulation capabilities (upper left). At the start of each scenario, the robot and the remote user Janet introduced themselves to the participant (upper right). We prepared three video vignettes to illustrate the remote experiences: visiting farmer's market (lower left); going to the botanical garden (lower middle); and visiting the lakefront (lower right). }
    \Description{The image is a visual summary of study materials used by study 1 session 2 involving a telepresence robot named Sparky and a user named Janet and the cover of three video vignettes. The top left panel shows an illustration of Sparky the Robot with interaction labels like "Likely," "Talk," and "Touch” and a note indicating a "Microphone". The top right panel includes a brief description of Janet, indicating that she recently went to the doctor and cannot go out for a while and Sparky will go to the Farmer's Market so Janet can still have a look around at home.The panel on the bottom shows three scenarios: a farmer's market stall displaying fresh vegetables, a botanical garden with a caption pointing out some pink flowers to observe and a lakefront with a caption that draws attention to two ducks visible in the distance.}
    \label{fig:design materials}
\end{figure*}

\paragraph{Scenario Design}
Guided by the insights from the initial interview, we developed three scenarios where the robot could provide a rich experience for homebound older adults: ``Visiting the Farmer's Market,'' ``Going to the Botanical Garden,'' and ``Visiting the Lakefront.'' For each scenario, we designed hypothetical interactions between a user and a robot in each scenario and created storyboards using video vignettes \cite{aguinis2014best}. To create a more relatable experience for the participants, we introduced the user persona, ``Janet,'' who is a homebound older adult and who sends the robot to explore the external world while she stays at home. Each video vignette was approximately two minutes long, and the premise of each scenario is outlined below:




\begin{itemize}
    \item \textbf{Visiting the Farmer's Market}---The robot walks through a crowded farmer's market, stopping at several vendors to explore what they are selling. The user gives periodic instructions to the robot about what she would like to do or see in the farmer's market. 
    \item \textbf{Going to the Botanical Garden}---The robot enters a botanical garden, then navigates through different sections in the garden. It also describes smells of flowers in the area and shows the plants' details such as water drops on the leaves.  
    \item \textbf{Visiting the Lakefront}---The robot travels along a walking path that follows a lakefront on a cold winter day. The robot describes different information to the user, such as the temperature of the day and different passersby sharing the path. The user asks the robot to go to the lakefront and have a close-up view of some ducks in the lake.  
\end{itemize}

\paragraph{Session 2: Interview using Video Vignettes}
In the second interview study, we used the video vignettes as stimulus to better probe older adults' experience with the conversational telepresence robot. We presented participants with an image of the robot that was annotated with its capabilities, including autonomous navigation and sensory functions such as seeing, hearing, touch and smell (Figure~\ref{fig:design materials}). 

After the introduction, we asked participants to choose one of the video vignettes based on their preferences. We watched the video together and then conducted a semi-structured interview to understand their feedback on the scenario and their overall preferences for the robot design and remote experiences. The interview questions focused on four areas: (1) Feedback for robot behaviors (\textit{e.g.}, guidance, control, and verbal description); (2) Preferences for social interactions through the robot; (3) Additional scenarios where the robot can be used; and (4) Challenges and concerns for using the robot. All participants viewed and discussed two to three of the video vignettes within the one-hour study duration. Audio and video recordings of each session were collected for analysis.


\subsubsection{Data Analysis}\label{study1_data_analysis}

The video recordings were transcribed and organized into spreadsheets. The transcripts were analyzed following the guidelines by \citet{braun2006using} for thematic analysis. 
Two coders worked together throughout the analysis process. Both coders were present for all study sessions, and thus were already familiar with the data. They first open coded two participants' transcripts. The initial codebook was created after consolidating the open codes and resolving all disagreements. Both coders independently coded the remaining data and updated the codebook through an iterative process of coding and discussion to ensure agreement. The codes were further grouped into clusters by linking related codes, and the final themes were created by affinity diagramming. 



\subsection{Findings}

We sought to understand the needs of homebound older adults in using a conversational telepresence robot to explore the external world. We found participants' desired remote experiences through the robot, \textit{i.e.,} reminisce, exploration and social participation. We also identified their expectations for the robot's capabilities and concerns when using the telepresence robot.

\subsubsection{Reminiscent experience}
Five participants (S1.P1, 3, 5, 7, 9) desired \textit{reminiscent experiences}, \textit{i.e.,} they indulged in past memories after seeing familiar scenes and wanted to go back to old places through the robot. 

\paragraph{Revisit Familiar Locations} 
Being homebound, participants had limited access to places that they used to work and live, but they expressed that they missed seeing them again. For example, P5 mentioned that he liked the ``\textit{nostalgic and warm feeling}'' (S1.P5) of the reminiscent experience after watching the video of the college campus he used to work at. Additionally, participants wanted the robot to show changes in places they used to know (S1.P1, 7, 9). For example, P1 wished to check the plants he used to know in the botanical garden scenario, ``\textit{to see if a certain plant or flower was in bloom that I, that I remembered has been there}'' (S1.P1). 


\paragraph{Recall the Past Memories and Remember the Disappearing}
Certain scenes and senses during the remote visit had participants recall past events. The botanical garden scenario reminded S1.P6 of her son's wedding because he ``\textit{got married over there.}'' The robot's description of gardenia smell reminded S1.P3 of their corsage experience in college. \rr{As S1.P3 shared: ``\textit{In the US, in the college dances, they use a lot of gardenias for, for corsage... I used to go the dances, and, not very serious, young men, but it was fun.}'' (S1.P3)}. Similarly, S1.P5 described how seeing the waterfront video vignette evoked her memories of how the lake smelled, saying that ``\textit{I can smell it}'' after watching the video. One participant (S1.P3) lost her taste from an illness and she wanted the robot to associate the taste of a dish with the dish she used to know by saying that ``\textit{This is what you use to like.}''(S1.P3) so she can recall the taste. Additionally, one participant (S1.P7) wanted the robot to record activities in a family farm that was going out of business,\textit{ e.g.}, ``\textit{go to all these different locations on the farm and get good pictures and information about it}'' (S1.P7).


\subsubsection{Exploratory experience}
Participants (S1.P1, 5--9) wanted to explore unusual or new things through the robot which we therefore refer to as \textit{exploratory experience}. \rr{They asked for information about plants' native habitats and growing conditions in the botanical garden (S1.P1, 5, 9), wanted the robot to identify buildings and wildlife at the lakeside (S1.P7), and wished to explore art museums and local fairs through the robot (S1.P3, 6). For example, S1.P3 wanted the robot to explain artwork in a museum and ask staff to get the related information, saying that: ``\textit{[the robot should] show you the art and then get whatever information they can for the, the people who work there. Some people would have nice information.}''} Additionally, S1.P6 and S1.P8 wanted the robot to look for \textit{specials of the day} and new foods that they had never tried before in the farmer's market visit. \rr{As S1.P6 shared: ``\textit{I'd always be looking for specials of the day or something that's new. I don't know. Anything different, I love to see different and new vegetables or or things that they had.}'' In addition to seeing and hearing about the remote environment, two participants (S1.P8, 9) felt that an important component of exploration included having the robot bring something back to them at home, such as fresh avocados and corn on the cob from the farmer's market.}





\subsubsection{Social participation}
Participants desired social experiences through active and passive social participation in the remote environment. Active social participation includes one-on-one conversations and interactions with friends and bystanders, while passive social participation refers to experiencing the social atmosphere and having a sense of belonging to community without direct interactions with people.

\paragraph{Active social participation} Three participants (S1.P5, 7, 8) wished to have one-on-one conversations and interactions with people who walked by in the remote environment. They thought it could be ``\textit{fun}'' (S1.P7) and wanted to say ``\textit{Hi}'' (S1.P5). \rr{S1.P8 wanted the robot to find a bystander to establish conversation with, saying that ``\textit{If you could see somebody, you know, and [the robot] could say, `So and so can you stop a minute and talk to [participant name]?'}'' S1.P7 also shared the same idea, saying that ``\textit{People who are at that environment. [Robot Name] will go up to them. And I can has [Robot Name] ask them certain things.}'' Notably, both participants (S1.P7, 8) expressed concerns about communication difficulties and wanted the robot help to convert the messages. For example, S1.P8 wanted the robot to ``\textit{transfer}'' the message and ``tell the person what I'm saying'' (P8). S1.P7 shared that she often had difficulty in communication when people ``\textit{talk too fast}'' or ``\textit{have an accent}'' (S1.P7). She explained that the robot could tell her what the person is trying to say if she could not hear clearly: ``\textit{Sparky [the robot] could do a better job than the people. How they talk. Maybe I get better information. Could hear it better}'' (S1.P7).}  

\paragraph{Passive social participation} Four participants (S1.P1, 6--8) reported how they enjoyed passive social participation such as the community gathering atmosphere and observing people in public space. \rr{For example, S1.P7 shared how she wished to observe the crowd ``\textit{at one corner},'' saying, ``\textit{See where that person is going. Just to get the whole ambience of the place. Could be of interest.}'' Similarly, S1.P6 described how she wanted the robot to ``\textit{wander}'' in a local fair: ``\textit{Like Fourth of July, wandering around and seeing people who hadn't seen for a while.}'' Additionally, S1.P1 mentioned how he wished to observe children playing at a playground and learn about the social dynamics among children and their parents during the play, sharing that: ``\textit{What are young children, what are they doing? Is it going down on slides and using equipment? Is that the primary thing? Or do they develop games of their own?}'' (S1.P1).} 





\subsubsection{Expected robot capabilities}

Our findings revealed a mixture of preferences for how participants wanted to interact with the robot, including controlling the robot through low-level controls and high-level instructions, and the desire for the robot's guidance and recommendations.


\paragraph{Low-level control of the robot} Low-level controls of the robot include controlling the robot's speed of navigation, speed of speech, focus of the view, and stopping the robot (S1.P1, 6, 7). Low-level controls allowed the user to see everything in the environment and manipulate the pace of the visit as they wished. The robot may ``\textit{focus}'' (S1.P7) on certain things and ``\textit{move past}'' (S1.P1) other things that may not be of interest to the user. The capability of robot control also protects the user's sense of agency and allows them to ``\textit{stay in charge}'' (S1.P6). When participants witnessed unexpected things, they could ask the robot to ``\textit{go back and see it again}'' (S1.P1). 

\paragraph{High-level instructions} The robot was also expected to follow higher-level instructions from the users (S1.P3, 4, 6, 8, 9) because the low-level robot controls were mentally demanding and users lacked knowledge of the environment. \rr{High-level instructions include giving the robot tasks ahead of time (S1.P6, 8) and asking the robot to provide guidance and make recommendations (S1.P3, 4, 6, 8, 9). For example, two participants (S1.P6, 8) preferred to delegate tasks to the robot, stating that ``\textit{I give [the robot] the list of things that I wanted}'' (S1.P6) and ``\textit{I'd want to tell the robot ahead of time that I'm wanting to get out there to get a certain thing}'' (S1.P8). Both participants shared that their lack of confidence in using the system, \textit{i.e.}, S1.P6 feared controlling the robot because she may ``\textit{push the wrong buttons}'' and S1.P8 mentioned that the robot could communicate better and ``\textit{probably be smarter}'' than her.} 

\rr{Additionally, five participants (S1.P3, 4, 6, 8, 9) mentioned that they desired the robot's guidance because of lacking the up-to-date knowledge of the environment. For example, S1.P1 commented that they did not know as much about the environment as the robot: ``\textit{I might not know the things that actually would be most interesting... But Sparky [the robot] would.}'' Similarly, S1.P3 explained the robot knows about the layout and what to focus on in the remote environment: ``\textit{Sparky [the robot] would know more about the layout of the land. What is available? What he should be focusing on?}'' Therefore, they desired the robot's guidance and recommendations. }

\subsubsection{Concerns for using the robot}
Participants were concerned whether or not bystanders would be socially accepting of the robot. For instance, S1.P8 mentioned that she did not want to be the center of attention and be the only person using the robot, \rr{as P8 stated \textit{``But if it was the only robot, going there, and stuff. I don't think I would feel comfortable.''}} S1.P6 expressed her concern for the robot's capabilities when it is not common out there, saying that ``\textit{Until it got to be a very common practice I'd be concerned all the time [Laugh]... What if she [the robot] is getting screwy or what?}'' (S1.P6). In addition, using the robot to explore a site that users previously visited may reinforce users' loss of capability. For example, S1.P6 mentioned that she used to enjoy hiking and seeing plants in the state park, but seeing the park through the robot can ``\textit{make me sad that I couldn't do it}'' (S1.P6). 

\subsection{Discussion}

Results from the needfinding study highlighted three desired experiences through the robot: reminiscent experience, exploratory experience, and social participation. \rr{Participants expected to interact with the robot through low-level control commands and high-level instructions, echoing the findings of \citet{tsui_accessible_2015}. Notably, participants expected the robot to take a guiding role and make recommendations based on their preferences due to their lack of knowledge of the environment and confidence in controlling the robot. Participants desired passive social participation in the remote environment through the telepresence robot, such as observing the crowds and experiencing the community atmosphere. This passive social participation is in addition to the robot's supporting communication with friends or families which have been widely studied in the prior work \cite{isabet2021social}.} Concerns for using the robot focused on the social acceptance of the robot in the environment as well as the risks of reinforcing decreased autonomy. The findings point to the following design implications: 


\begin{itemize}
    \item \textbf{Interactive dialogue to curate user experience.} 
    Given the user needs of exploring and learning in the remote experience, the robot can provide guidance and respond to the user's spontaneous queries about the environmental information. The robot can use narratives to describe smells, temperature, and atmosphere in the environment to supplement the audio and visual information from the camera's live streaming. 
    
    \item \textbf{Facilitate meaning making.}
    Reminiscing is a central theme within our findings. Homebound older adults desired to visit their old neighborhoods or places with significant meanings. Actively listening to and responding to homebound older adults when they share their past during the experience can facilitate their meaning-making process and improve their sense of companionship. 
   
    \item \textbf{Mediate social interactions with bystanders.} Communication through the robot can be challenging for both the remote user and local people due to hearing difficulties, lack of technology proficiency, and environmental noises. To support users' social participation in the remote environment, the robot can initiate social interactions on the users' behalf. The robot can also transfer the message if the environmental noise is high so that the remote user can more easily understand.
\end{itemize}

Through a series of video vignettes, the needfinding study revealed desired user experiences through the robot and users' expectation of the robot's capabilities. However, the experiences of participants with the presented scenarios may differ from those with functional prototypes given the barriers to using technology they already have. Therefore, we prototyped a conversational telepresence robot following the expected robot capabilities and design implications from the needfindings study. Then we conducted the second study to understand user experiences with and responses to this robot prototype.


\section{Study 2: Understanding Preferences and Perceptions Using a Technology Probe}
In the second study, we aim to answer the following research question: what might be their experience with exploring and interacting with the external world via a conversational telepresence robot? In particular, we aim to understand older adults' interaction patterns in their desired experiences. 



\subsection{Method}
 \begin{table*}[!t]
    \caption{Demographic information for our participants in the Technology Probe study. We derived the scales for mental health, social connection, and technology proficiency from the onboarding questionnaire items. The scales have a range from 1--5 with 5 indicating the highest score.}
    \label{tab:study 2 demographics}
    \centering
    \renewcommand{\arraystretch}{1.2}
    \small
    \begin{tabular}{p{0.025\linewidth}p{0.035\linewidth}p{0.065\linewidth}p{0.12\linewidth}p{0.1\linewidth}p{0.15\linewidth}p{0.075\linewidth}p{0.1\linewidth}p{0.1\linewidth}}
    \toprule
        \textbf{ID} & \textbf{Age} & \textbf{Gender} & \textbf{Time living in the care facility} & \textbf{Mobility Aids} & \textbf{Days Out Per Week} & \textbf{Mental Health} & \textbf{Social Connections}  & \textbf{Tech. Proficiency} \\ 
        \midrule
        S2.P1 & 96 & Female & 2.5 years & Walker & 2 & 4.5 & 3 & 5\\
        S2.P2 & 92 & Female & 0.5 year & None & 2 & 3.25 & 4.5 & 3\\
        S2.P3 & 91 & Female  & 1 year & Walker & 0 & 4.75 & 3.5 & 3\\
        S2.P4 & 82 & Female & 3 years & Walker & 0 & 4.75 & 5 & 5\\
        S2.P5 & 92 & Female & 1 year & Walker & 7 (sitting outside), 1 (visiting other places) & 4.5 & 4.5 & 1\\
        S2.P6 & 86 & Female  & 3 years & Cane & 7 (sitting outside), 0-1 (visiting other places) & 3.75 & 2 & 4.5\\
        S2.P7 & 95 & Female  & 2 years & Cane, Walker & 7 (sitting outside) & 4.5 & 5 & 4.5\\
        S2.P8 & 97 & Female  & 5 months & Walker & 2-3 (sitting outside), 0-1 (visiting other places) & 3.25 & 4 & 4.5\\
        S2.P9 & 98 & Male  & several years & Walker & 3 (sitting outside) & 3.5 & 3 & 3\\
        S2.P10 & 81 & Female  & 7.5 years & Cane & 7 (senior transportation service needed) & 4 & 4 & 3\\
        S2.P11 & 80 & Female  & 3 months & None & 7 (sitting outside), 1 (visiting other places) & 4.5 & 5 & 1\\
        S2.P12 & 83 & Male  & 1.5 years & Walker, Wheelchair & 0 & 5 & 5 & 5\\
        \bottomrule
    \end{tabular}
\end{table*}

\subsubsection{Participants}
We recruited 12 participants (10 females, two males) aged 80--98 ($M=89.42$, $SD=6.67$) from a senior living facility located in the Northeastern United States. One participant (P7) dropped out from the study after the onboarding session because of medical reasons. Nine participants reported using at least one mobility aid, \textit{i.e.}, cane, walker and wheelchair, and the two participants who did not use mobility aids reported leaving their dwelling environment once or twice a week. Therefore, all participants were considered to be homebound. Participants' demographic information was reported in Table \ref{tab:study 2 demographics}.

\begin{figure*}[]
    \centering
    \includegraphics[width=\textwidth]{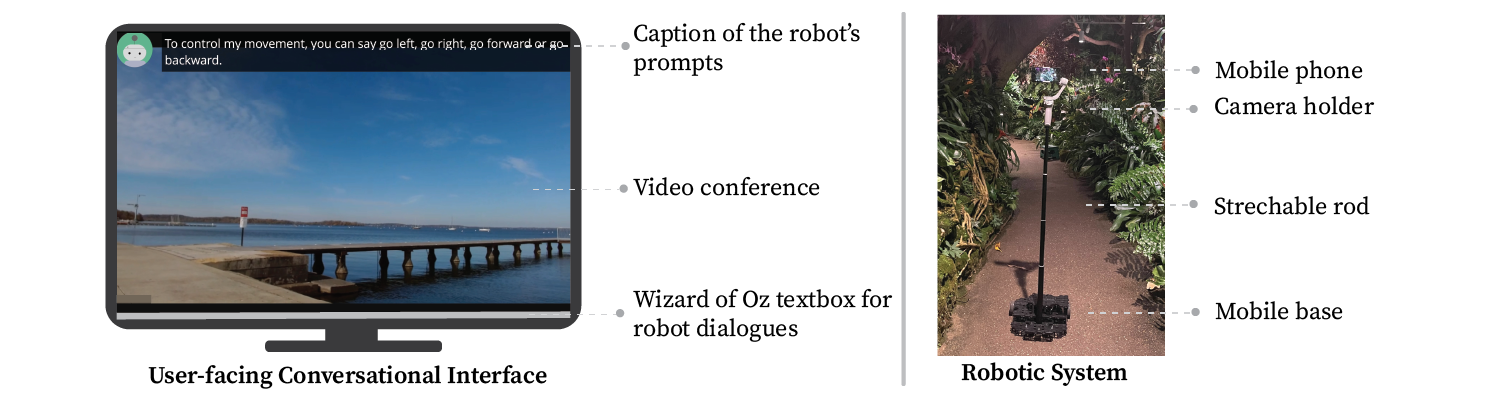}
    \caption{Components of the \textit{Technology Probe} study interface we designed.}
    \Description{Set up of the Technology Probe study interface. The left side of the figure illustrates an example of the user-end conversational interface, which contains an icon of the robot and the captions of robot’s on the top and a serene view of the lakefront at the bottom. The robot says, "To control my movement, you can say 'go left,' 'go right,' 'go forward' or 'go backward.'" The right side of the figure shows the component of the robot which include a Mobile phone mounted on a Camera holder on the top, and attached to a Stretchable rod that is connected to a Mobile base, enabling movement and interaction within the environment.}
    \label{fig:robot}
\end{figure*}

\subsubsection{Apparatus} 
We prototyped a conversational telepresence robot using a Turtle bot mobile base, a stretchable rod, and a camera holder with a mobile phone on the top (Figure \ref{fig:robot}). Users interacted with the robot through a conversational interface which we designed based on the findings and implications from the needfinding study. A researcher remotely controlled the robot's navigation and dialogue through a Wizard-of-Oz interface according to the user's verbal inputs to reduce the technical complexity and prevent usability issues from affecting participants' experience. 

The conversational interface was built as an overlay on top of a video conferencing interface (Figure \ref{fig:robot}). The video conferencing technology utilized the existing commercial products (\textit{i.e.}, Zoom \footnote{Zoom: \url{https://zoom.us}} and Discord \footnote{Discord: \url{https://discord.com/}}). The conversational interface was implemented as a Chrome extension and the robot responses had two sources: pre-scripted prompts triggered by keyboard shortcuts and free text inputs typed by the experimenter. The robot dialogue has three parts. First, it provides guided narrations about the remote environment. Second, it provides social chat with the user and supports open-ended questions. Third, it facilitates the user's communication with the bystanders and conveys the message for the user if the communication failed.


\begin{figure*}[!b]
    \centering
    \includegraphics[width=\linewidth]{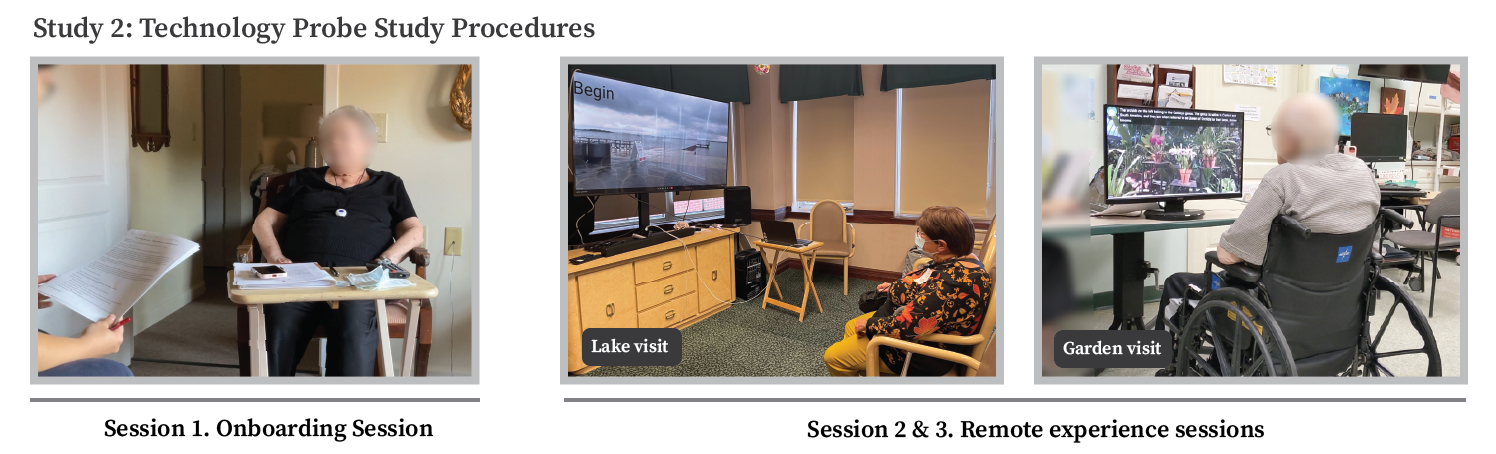}
    \caption{Overview of the \textit{Technology Probe} study. We first held on boarding sessions with older adults and filled in questionnaires about their physical and mental health conditions and experiences with technology (left). In the second and third sessions, participants remotely visited a lakefront park next to a university campus and a botanical garden through the robot where they engaged in three phases: free exploration, chatting with the robot and chatting with the remote bystander (right).}
    \Description{The figure presents the procedure of the technology probe study. The left picture shows the onboarding session, where an older adult is sitting at home and answering questions from an experimenter. The left two pictures demonstrate the remote experience sessions. In the first picture on the left, a female older adult is watching the lake view on TV livestreamed from the robot. In the second picture on the left, a male older adult is watching the botanical garden view on a monitor livestreamed from the robot. }
    \label{fig:method}
\end{figure*}

\subsubsection{Study Procedure}
The technology probe study consisted of three sessions for each participant. The onboarding session served to establish rapport and assess the participant's health conditions and experience with technologies. The second session and third sessions included experiencing at a lakefront park and a botanical garden through the conversational telepresence robot. The sessions took place either in the participant's room or in the community area in the senior living facility and all sessions were video and audio recorded. All the study materials including the consent form, study protocol, and surveys were approved by the Institutional Review Board (IRB) at the University of Wisconsin–Madison. 

In the onboarding session, the experimenter conducted semi-structured interviews that asked participants about their day-to-day experiences. Then participants completed a survey about their physical health and mental health conditions, and their experience with technology by writing on a printed form (one participant) or providing answers verbally as the experimenter read the survey items aloud (11 participants). At the end of the session, we scheduled the time for the next session for the participant to remotely visit the lakefront or the garden through the robot. 

In the second and third sessions, the participant remotely visited a lakefront by a university campus and/or visited an indoor botanical garden. All participants were offered to remotely visit both locations, but six participants only visited one location because of their personal preferences or their availability for the visit. In total, four participants opted for the lake visit, two participants opted for the garden visit, and five participants opted for both visits . 

For each visit, participants first had a training phase to gain familiarity with the verbal interaction with the robot. During the training, participants were told that they can verbally control the robot or ask questions to the robot. Then the experimenter asked them to try a few verbal commands such as ``turn left'' or ``turn right'' to see the view change because of the robot's movement.   

After the training phase, participants were told that they would have a three-phase experience with short interviews in-between the phases: (1) Exploration phase, (2) Robot chat phase, and (3) Bystander engagement phase. 

\paragraph{Exploration phase} Participants first engaged with the robot in free exploration of the remote environment. The robot first had a self-introduction, saying ``\textit{Hello, this is Jackie the Robot. We will take a walk at the botanical garden/lakefront today.}'' Then the robot asked the participant for their preferences for guidance or control: `\textit{`Do you want me to guide you, or do you want to control it yourself?}'' Each location contained multiple points of interests for the robot to provide guidance. When the robot navigated near each point of interest, the robot asked the participant again whether they would like the robot to guide or control on their own. After visiting three areas of interests, the experimenter stopped the robot and conducted a semi-structured interview. Participants were asked about their preferences for the robot's guidance and control, whether they felt in control of the robot and in control of what they wanted to do, as well as their impression of the robot's narrations and guidance.

\paragraph{Robot chat phase} In the second phase, the robot initialized social chat with the participant and asked questions about their personal past. For example, during the visit at the lakefront next to a college campus, the robot asked ``\textit{Are there any special memories when you were a student?}'' The robot followed up with the participant until they stopped proactively engaging in the conversation and then asked another triggering question. After all of the questions were asked, the experimenter stopped the robot again and asked the participant to reflect on the experience with a semi-structured interview. Participants provided feedback on the conversation with the robot and other things they would like to chat about with the robot.

\paragraph{Bystander engagement phase} In the third phase, the robot initiated a conversation between the participant and a remote bystander. The bystander was an experimenter who pretended to be a visitor in the remote environment. Participants were not told that the bystander was from the research team to make the experience as realistic as possible. First, the robot asked the participant, ``\textit{There is a visitor walking to us and seems interested in the robot. Do you want to say Hi?}'' If the participant agreed to talk to the bystander, the remote experimenter would start chatting with the participant through the robot, asking questions such as ``\textit{How are you?}'' ``\textit{How is your visit today?}'' During the interaction, the robot would ask the participant if they would like to show their face to the bystander. After the third phase, participants were asked about their feedback on the bystander interaction.

At the end of each visit, we interviewed participants about their overall experience, feedback for talking to the robot, challenges in the interaction, and preferences for other people joining the experience together with them. The onboarding survey, semi-structured interview questions, pre-scripted questions from the robot and the examples of the dialogue is documented in the Appendix in the osf repository \footnote{OSF repository: \url{https://osf.io/eq3zn/?view_only=48a1c7bf3a284f75bde8331a067f7b62}}.


\subsubsection{Data Analysis}

All sessions were first transcribed through an automatic transcribing service and then manually verified by three researchers on the team to correct any errors from the automatic transcription. The transcriptions were first categorized by the phases in the study, \textit{i.e.}, the exploration phase, robot chat phase, and bystander chat phase. We analyzed the transcriptions based on a thematic analysis approach \cite{braun2006using} following the same approach from the needfinding study (see \S\ref{study1_data_analysis}). Our goal was to understand older adults' interaction patterns in their desired experiences, so we grouped codes into clusters representing the high-level themes. Under each high-level theme, we generated the sub-themes by further linking the related codes through affinity diagramming. 


\begin{figure*}[]
    \centering
    \includegraphics[width=\textwidth]{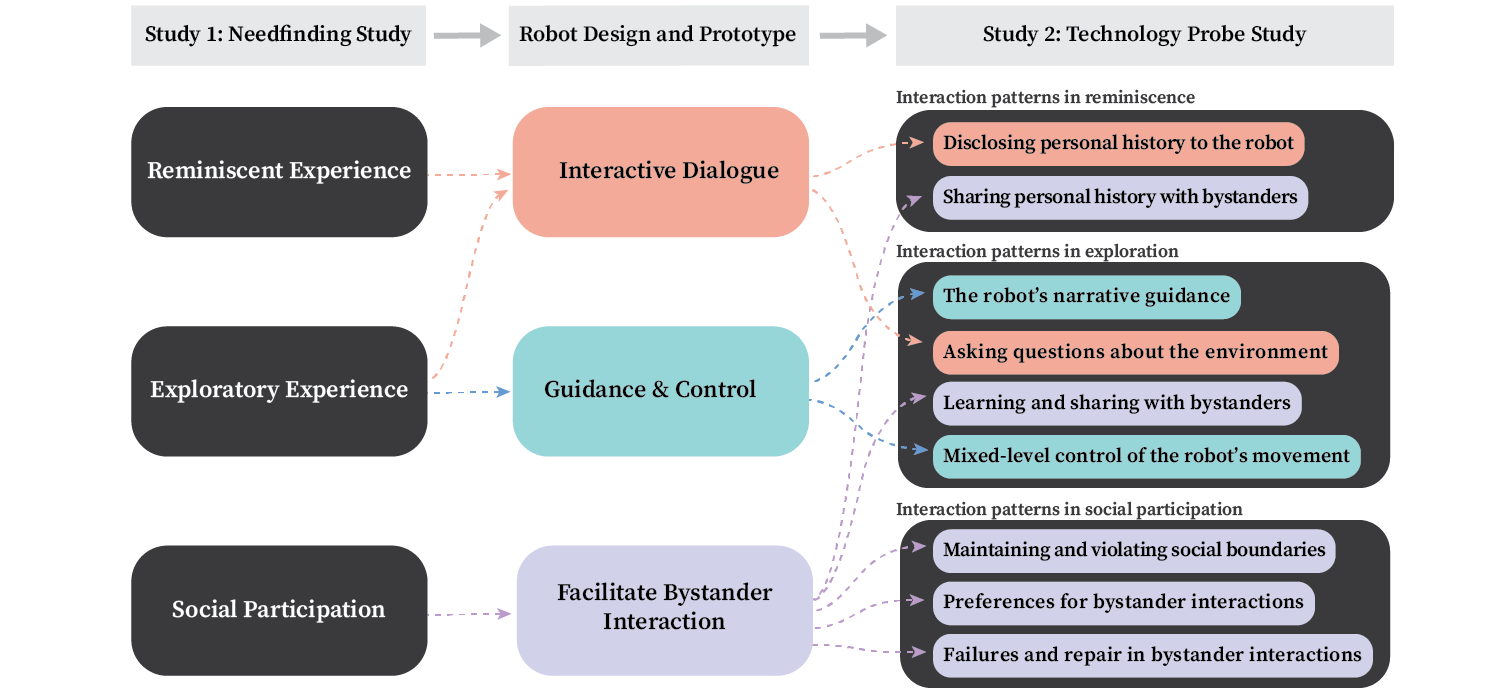}
    \caption{Finding overview from our needfinding study and technology probe study. We first conducted a needfinding and identified homebound older adults' desired experiences in the telepresence experiences (Left); Guided by the findings, we prototyped a conversational telepresence robot which supports interactive dialogue with the user, provides guidance and control, and facilitates bystander interaction (Middle). We conducted a technology probe study and identified interaction patterns in each experience between our participants and the telepresence robot (Right).}
    \Description{The figure is a flowchart divided into three main sections, including the needfinding study, robot design prototype, and technology probe study. The first section on the left lists three key experiences:Reminiscent Experience, Exploratory Experience, Social Participation. The central section in coral and teal colors identifies the three main functions of the robot: Interactive Dialogue, Guidance \& Control, Facilitate Bystander Interaction. Dashed red lines connect the experiences from the finding study to the corresponding functions in the robot design, where Reminiscent Experience and Exploratory Experience are both connected to Interactive Dialogue, Exploratory Experience is connected to Guidance \& Control, and Social Participation is connected to Facilitate Bystander Interaction. The final section in the flowchart contains interaction patterns in three areas: Interaction patterns in reminiscence which includes Disclosing personal history to the robot, and Sharing personal history with bystanders; Interaction patterns in exploration section which contains The robot’s narrative guidance, Asking the robot questions about the environment, Learning and sharing with bystanders,and Mixed-level control of the robot’s movement; Interaction patterns in the social participation section which contains Maintaining and violating social boundaries with bystanders, Preferences for bystander interactions, Failures and repair in bystander interactions. The image is a flowchart divided into three main sections, outlining the progression from a finding study to robot design and prototype, and then to a technology probe study. The lines indicate that Interactive Dialogue is examined through interaction patterns in reminiscence, such as disclosing personal history to the robot or bystanders. 'Guidance \& Control' is evaluated through interaction patterns in exploration, including guiding narratives, environmental inquiries, and movement control. 'Facilitate Bystander Interaction' is tested through patterns in social participation, which involve maintaining social boundaries, preferences for bystander interactions, and Failures and repairs in bystander interaction.}
    \label{fig:theory}
\end{figure*}

\subsection{Findings}

Our findings reported the user preferences for the experience, interaction patterns in the reminiscent experience, exploratory experience and social participation, all participants favored the robot's guidance over controlling in the robot in the remote experience due to their expectation of the robot's guidance role and various challenges in controlling the robot. In the end, we reported participants' overall experiences, including their overall positive feedback and barriers in the interaction.



\subsubsection{\textbf{Interaction Patterns in Reminiscent Experiences}} \label{sec: interaction patterns in reminisce}

Three participants (S2.P2, 9, 11) reminisced about their past experiences when seeing familiar scenes in the remote environment. This reminiscent experience was facilitated by sharing and disclosing their personal experience to both the robot and the bystander. 

\paragraph{\textbf{Disclosing personal history to the robot}}
Participants' dialogue with the robot supported reminiscent experiences, as they shared with the robot their past experiences with family and friends at gardens, lakes, and colleges (S2.P2, 9, 11). For example, after seeing the orchids in the botanical garden, S2.P11 shared with the robot a story about when they received orchids as gift from their neighbors. Later, when the robot asked about their favorite plants, S2.P11 shared another story when a rabbit ate all the tulips under her daughter's window.


Watching the lake reminded S2.P2 of her husband who just passed away. In response to this bittersweet memory, she wanted to name the robot after her late husband so she could have the experience as if they were enjoying the view together. As S2.P2 described: 
\begin{quote}
    \textbf{S2.P2}: So hearing robot's name is [her husband's name] would have been nice. That's [her husband's name] and I again, enjoying the things we enjoy together.
\end{quote}

S2.P2 further shared with the robot about trips she had with her husband, saying that ``\textit{This scene reminds me this beautiful scene of the trips, we used to go in the south to different places that had water.}'' She told the robot that she wished to be on the boat on the horizon because she use to do that with her husband. As she described: ``\textit{I see boats in your background. I think I've always been a sailor at heart and I like to be on that, on that boat [stutter] with my husband [name], and we did many times.}'' (S2.P2).

Since the lakefront was located by a college campus, two participants (S2.P2, 9) recalled memories of their experiences as students from decades earlier. S2.P2 shared the story with the robot that she wasn't able to attend college when she was younger because of the cost and returned to school when she was forty: ``\textit{I went back to school when I was forty, not when I was very young, because I could not afford to go to school. I had to work my way through college.}'' (S2.P2). When asked what could be improved when interacting with the robot, S2.P9 stated how he wanted to share ``\textit{an experience that I had on campus that was really significant}'' and he could provide more ``\textit{comprehensive}'' answers to the robot if the conversation continued (S2.P9). 

However, two participants (S2.P5, 6) reported the discomfort sharing with strangers when the robot asked personal questions. For example, S2.P5 did not want to talk about themselves and commented that that ``\textit{The robot doesn't know me. I don't know the robot.}'' Similarly, S2.P6 stated that ``\textit{I don't have any connection to this person. We have nothing, as far as I know, to talk about.}''

\paragraph{\textbf{Sharing personal history with bystanders}}
\label{sec: chat with bystander for reminisce}
In addition to conversational exchanges with the robot, participants chatted with bystanders in the remote environment. During the lake visit, S2.P2 asked the bystander, who was a undergraduate student in mathematics from the research team, about her class and her major. S2.P2 mentioned that she ``\textit{could relate}'' with the student's experience and compared it with her past situation as a female student in science ``\textit{back in the 60s}:'' 

 \begin{quote}
         \textbf{S2.P2}: ``That's almost like me when I was young, I was interested in mathematics, but I was not permitted to take chemistry when I was 16 years old, because women were not permitted to take sciences in the United States. And I had to go into a profession that I never really liked.''
 \end{quote}

Notably, engaging with the bystander played a major role in the experience for this participant (S2.P2). When asked to compare the lake visit and the garden visit, S2.P2 preferred the lakefront visit because she had talked to the bystander about their school experience. As she commented: ``\textit{I was so happy to see that young women are interested in the sciences. I could relate because I was too}.'' (S2.P2).

\subsubsection{\textbf{Interaction Patterns in Exploratory Experiences}}
\label{sec: interaction patterns in exploration}
The majority of participants reported that their main usage goals of the robot were to learn and to explore (S2.P1, 2, 3, 5, 8--11). The robot was seen to ``\textit{help with my curiosity}'' (S2.P2), the experience was``\textit{educational}'' (S2.P8) and ``\textit{there are always things that you learn when you go each time}'' (S2.P3). The exploratory experience was achieved through the dialogue with the robot and interactions with the remote bystanders. Below we report participants' interaction patterns in the exploratory experience.


\paragraph{\textbf{The robot's narrative guidance}} The primary ways of learning about the environmental information was through the robot's narrations when participants asked it to guide the visit. Participants who held positive feedback for the robot narrations (S2.P1-3, 9-12) described them as ``\textit{informative}'' (S2.P10, 12), ``\textit{specific}'' (S2.P2), and ``\textit{helpful}'' (S2.P9). They expected the robot to guide and have knowledge about the environment (S2.P2, 3, 11, 12). As S2.P12 commented: ``\textit{It's nice to be led by particularly because the robot has the captions and knows what we're looking at.}''

Participants also pointed out areas of improvement for the robot's narrations or disliked the narrations (S2.P1, 3, 5, 6, 8). Two participants wished that the narrations could be more ``\textit{professional}'' (S2.P8) and provide ``\textit{more details}''(S2.P3). For example, S2.P8 found the narratives ``\textit{amateurish}'' and wished to hear things that benefit ``\textit{educationally, emotionally, technologically}'' (S2.P8). Two participants(S2.P1, 8) disliked the robot's narrations because they were uninterested in the location. For example, S2.P1 was uninterested in orchids and thought the narrations were not exciting, stating that ``\textit{I just learned something about orchids but I'm really not interested in orchids.}'' Another participant (S2.P6) did not like the robot's ``\textit{enuciation}'' and thought the robot was ``\textit{just blabbering away behind some pictures.}''

\paragraph{\textbf{Asking the robot questions about the environment}}
Five participants (S2.P2, 3, 4, 9, 11) proactively engaged in the dialogue with the robot and asked questions about the environment. During the visit at the lakefront, participants asked for the geographical information about the lake (S2.P2, 3, 4, 9), water sports (S2.P3), the buildings and their functions (S2.P2, 8, 10) and student life on campus (S2.P2, 9, 10). \rr{For example, during the lake visit, S2.P9 wanted the robot to look for the buoy in the lake after the robot's introduction, asking: ``\textit{Where is the buoy? What does it look like? And is it under the water?}''} When visiting the botanical garden, participants wanted to learn about the plant name (S2.P2, P11), flower types and origin (S2.P3), and their lifespan (S2.P10). Participants expected that the robot to have knowledge about the environment and answer their questions (S2.P2, 3, 11). As S2.P11 shared, ``\textit{the robot knew the answer all the time.}'' One participant (S2.P3) even preferred talking to the robot than talking with people, because ``\textit{the robot has the answers a lot faster than persons}'' (S2.P3).


    

     


\paragraph{\textbf{Learning and sharing with bystanders}}
\label{sec: chat with bystander for exploration}
In addition to asking detailed questions to the robot, participants chatted with the bystander to learn new information and perspectives for the environment (S2.P2, 3, 9, 11). For example, S2.P9 wanted to share their experience with the remote bystander because ``\textit{different people see different things}'' and ``\textit{it's really worthwhile educationally to get another impression of what it is we're both seeing}'' (S2.P9). S2.P2 thought that the bystanders can point out areas they ``\textit{may have missed}'' (S2.P2). 

\paragraph{\textbf{Mixed-level control of the robot's movement}} All participants had chosen the robot's guidance at least once when asked if they would like the robot to guide them or control on their own during the exploration phase. Mixed-level control commands were also observed during the interaction where participants used low-level control commands (S2.P4, 9) and instructed the robot to go to specific locations (S2.P9, 10). For example, in the botanical garden, S2.P9 controlled the robot to move close to and far away from flowers to see them from different angles \rr{with commands in the following sequence:``\textit{Go right. Stop. Now go left. The flowers are lovely. Stop. Continue going left. Stop. Go right. Stop. Now go forward. Now go backward. Stop.}'' This participant (S2.P9) shared how he was able to see the flowers in detail by controlling the robot's with these commands: ``\textit{It gave me the opportunity to really get a clearer picture of the purple of orchids and the white ones.}'' (S2.P9).}

\subsubsection{\textbf{Interaction Patterns in Social Participation}}
In the previous sections, our findings showed social participation facilitated the reminiscent experience (\S\ref{sec: chat with bystander for reminisce}) and exploratory experience (\S\ref{sec: chat with bystander for exploration}). Below we reported the general interaction patterns observed in the bystander engagement and users' feedback for this experience.

\paragraph{\textbf{Maintaining and violating social boundaries with bystanders}}
 All participants greeted the bystander after the robot initiated the interaction. One participant (S2.P2) cared about the friendliness of the bystander and asked the robot ``\textit{Is she smiling at me?}'' Although the majority of our participants' interactions with the bystander were polite, some cases of breaking the social norms were observed (S2.P1, 11, 12). Two participants (S2.P11, 12) asked personal questions to the bystander which could overstep personal boundaries (especially in the case that the bystander was not a research team member). For instance, P11 asked the bystander: ``\textit{Do you have children?}'' and S2.P12 joked with the bystander: ``\textit{Will you marry me?}'' One participant (S2.P1) exhibited impolite behaviors and said: ``\textit{No. Get out of the way.}'' after the robot asked if there is anything else she wanted to say to the bystander.

\paragraph{\textbf{Preferences for bystander interactions}} Four participants liked the bystander interaction through the robot (S2.P2, 10, 11, 12) and four participants provided negative feedback (S2.P1, 4, 5, 6). One participant (S2.P3) was positive towards the bystander engagement at the lakefront, but did not want to engage in the botanical garden session because she wanted to learn about the plants from the robot or an expert rather than the visitor bystander. Furthermore, three participants (S2.P1, 4, 6) reported that they did not want to talk to the bystander because that they did not know them. S2.P1 shared that ``\textit{I don't want to talk to strangers.}'' and S2.P4 felt ``\textit{odd}'' to chat with the bystander.

\paragraph{\textbf{Failures and repair in bystander interactions}}
Interaction failure often occurred where participants could not hear the bystander because of environmental noises and quality of the audio. This led to challenges in turn-taking, \textit{e.g.}, one participant (S2.P2) repeatedly asked questions without giving the bystander an opportunity to respond. Participants were able to continue the interaction after the robot conveyed the message for them. For example, when S2.P1 did not hear the bystander well, the robot conveyed: ``\textit{She was asking what's happening here.}'' And the participant continued the conversation by saying ``\textit{I'm being guided by with a robot.}'' (S2.P1).

\subsubsection{\textbf{Overall user experience of the novel system}}

We observed a gap in the acceptance of the telepresence experience among our participants. Five out of 11 participants (S2.P2, 3, 10, 11, 12) provided positive feedback for the experience, commented that they ``\textit{liked it very much}'' (S2.P2, 3, 11) and thought it was ``\textit{wonderful}'' (S2.P3, 11), ``\textit{fun}'' (S2.P12), and ``\textit{fascinating}'' (S2.P11). Five participants (S2.P1, 4, 5, 6, 8) disliked the experience and four of them even refused to continue the study after the first session visiting the lakefront. One participant provided mixed feedback for the experience.

\paragraph{\textbf{Positive experiences with robot-guided remote exploration}}

Below, we summarize the findings related to participants' overall positive feedback about the remote experiences.

\begin{itemize}
    \item \textbf{Access to the external world}. Participants reflected on how the robot helped them access places that they were not able to go or talk to people that they normally would not talk to (S2.P1, 2, 8, 12). For example, S2.P12 shared that he ``\textit{can't walk}'' and thought that this robot could take him to ``\textit{the botanical garden in Beijing or botanical garden in Jerusalem}.'' As he commented: ``\textit{This really lets us see the entire world.}'' (S2.P12). Two participants (S2.P1, 8) shared that this experience enabled them to see things or places that they ``\textit{would not normally see}'' (S2.P8). Also S2.P2 mentioned that ``\textit{The robot helped me talk to a young person that probably had no desire to speak to an older woman.}''



    
    \item \textbf{Personalized experience}. Three participants (S2.P2, 11, 12) shared that the telepresence experience was catered for their interests and desired pace. They appreciated that the robot ``\textit{pays attention}'' (S2.P11) and  ``\textit{was interested in me, personally'}' (S2.P2). S2.P12 compared this experience with documentary on TV and thought that ``\textit{the experience with the robot is more personal than just the documentary}'' (S2.P12).
    
    \item \textbf{Immersive experience}. Two participants shared that the robot created an immersive experience for them (S2.P2, 3). They commented that ``\textit{I feel like I'm there}'' (S2.P3) and ``\textit{This one was going along with me}'' (S2.P2). 

    \item \textbf{Interactive experience}.
    Participants saw the communication with the robot supported their ``\textit{curiosity}'' (S2.P2) and made a difference from watching a documentary. For example, S2.P9 commented: ``\textit{This is different because I can interact with the robot, but when I'm listening to the radio, if the radio is telling me something, I can't talk to the radio.}'' The interactivity also contributes to higher perceived agency of our participants (S2.P3, 12) because they had control over what they wanted to do. As S2.P3 shared ``\textit{I am doing it and not, not someone on TV.}'' S2.P12 also commented that ``\textit{We certainly do have the option to tell a robot to stop and spending more time in front of this flower or that flower.}''


    \item \textbf{Overcome accessibility challenges.}
    Having the robot could help to prevent the tiredness and overcome the mobility challenges (S2.P3, 11, 12). S2.P11 shared that there was one spot she used to like in the botanical garden but she could not go there often to see that because of the walk. However, she viewed the robot as one way to overcome this barrier. As she shared, ``\textit{Instead of my walking back and forth, [the robot] does it for me.}'' (S2.P11). S2.P3 also shared that ``\textit{If I was walking like I normally would be, then I would be tired.}'' 
    

\end{itemize}

\paragraph{\textbf{Negative experiences with robot-guided remote exploration.}} 
Overall participants reported the following challenges in interacting the robot: 
\begin{itemize}

\item 
\textbf{Lack of environmental awareness.} Most of homebound older adults in our study reported that they opted to have the robot guide their experience because of the lack of the environmental information (S2.P1, 2, 3, 6, 8, 9, 10, 12). Participants mentioned that they ``\textit{don't know where to go}'' (S2.P3) and ``\textit{don't know the area}'' (S2.P8). As a result, they viewed the robot as a guide and expected that the robot has knowledge of the environment. As S2.P12 shared: ``\textit{Presumably the robot knows where it's going. I don't know anything about that at all, so I'm perfectly happy to have a guide.}''

    \item 
\textbf{Lack of confidence in controlling.} Participants commented that they preferred the robot to guide them due to lack of knowledge or confidence to control the robot (S2.P3, 5, 6, 9). Two participants (S2.P3, 9) felt that they need more time to adapt to controlling the robot and become familiar with the environment. 





\item 
\textbf{Perceived opposite moving directions from the verbal commands}. Two participants were confused about the direction of the robot movement (S2.P2, 9) that they perceived the robot turning in the opposite direction from their control commands. They took the screen as the frame of reference rather than the remote robot. As S2.P2 commented: ``\textit{Turn left is this way. It's not this way. On my left or the label?}'' (S2.P2) 
\item 
\textbf{Lack of robot presence.} Four participants expressed confusion towards the robot's presence (S2.P5, 6, 8, 11) and the lack of robot presence can cause challenges in control and discomfort in the interaction. As S2.P5 shared: ``\textit{I don't feel anything. I don't have any control. Where is the robot?}'' 

\item
\textbf{Challenges in the verbal interaction.}
Four participants provided negative feedback in the verbal interaction with the robot (S2.P4, 5, 6, 8). S2.P5 felt uncomfortable talking to the robot and commented that ``\textit{I don't really, don't want to talk to the robot.}'' Participants also commented that they did not know what to ask, stating that ``\textit{I don't know what to ask.}'' (S2.P5) and ``\textit{I don't know all the choices that I have.}'' (S2.P8).





\item
\textbf{Difficulty in comprehending the experience.} We observed how difficulty comprehending the experience caused barriers in the experience. Three participants (S2.P4, 5, 6) reported difficulty in understanding the interaction and found the experience ``\textit{obscure}'' (S2.P5) and ``\textit{weird}'' (S2.P4). They thought it is challenging to adopt to new technologies because of the age gap and preferred things in the old way (S2.P4). S2.P5 mentioned that the younger generation has their ``\textit{mind}'' which she did not understand and she was ``\textit{completely at a loss}'' during the experience (S2.P5). S2.P6 felt frustrated when the interaction failed, saying that ``\textit{I don't understand what it's about. I don't find it very interesting, and I would like to know what the hell it's for}'' (S2.P6).


\end{itemize}

\section{General Discussion}
\rr{Our work uses a \textit{research through design (RtD)} approach to understand older adults' needs for remote experiences, resulting in a technology artifact which we designed and design knowledge generated from the technology probe study. \textit{RtD} has been widely used for digitally marginalized groups such as older adults to bridge the gap in design knowledge between typical users and older adults \cite{lindsay2012engaging, harrington2022s}. Our two studies illustrate an iterative process of user understanding and system design and refinement in \textit{RtD} \cite{lee2022unboxing, lee2023demonstrating}. In particular, our first study reveals the desired experiences by our participants through the discussion of our video vignettes. Study two uncovers more specific factors related to the usage and acceptance of the system, including participants' strong preferences for the robot's guidance, having dialogue with both the robot and bystanders for reminiscent experiences, communication challenges, and the violation of the social boundaries. }

\rr{Our findings highlight that having dialogue with the telepresence robot and bystanders fostered users' reminiscence and exploration in the remote environment and emphasize the need for the robot to take a proactive role in mediating the experience. Dialogue with the robot enabled users to have situated learning experiences where they can learn about objects in view and the social and historical information about the remote environment through the robot's narratives and inquiries with the robot. While prior work revealed the need to have autonomous and semi-autonomous navigation capabilities by telepresence robots \cite{tsui2011exploring, tsui_accessible_2015}, our findings highlight the importance of the robot's agency in providing guidance and making recommendations based on the users' preferences to create personalized experiences. Lastly, our findings emphasize the needs of social engagement in the remote environment, including both active and passive social participation.It was important to not only provide opportunities for meaningful conversations between the user and people in the remote environment, but also immerse the remote user in the social dynamics in the remote environment and reduce the sense of isolation. Overall, our work presents a novel conversational telerepsence system that opens opportunities for homebound older adults to access remote locations and create personal meanings through the interaction with the robot and the environment. This experience contributes to the independent living of older adults to engage in leisure activities even while remaining in their homes \cite{lazar_successful_2017, zhao2023older, zhao2024older} and to enhancing the their psychological wellbeing \cite{adams_critical_2011, morse_creativity_2021}. 
} 










\subsection{Design Implications}

\rr{Below, we synthesize findings from both studies to present the design implications for the three desired experiences, \textit{i.e.}, exploratory experiences, reminiscent experiences, and social participation. }












\subsubsection{Design Implications for Exploratory Experience}
\rr{Older adults desire to use telepresence robots for remote experiences such as going to museums, concerts and sporting events \cite{beer_mobile_2011}. In this work, we advance the knowledge of how older adults might interact with telepresence robots in leisure and exploratory activities through a conversational interface. Specifically, our findings highlight the importance of the robot's proactive role in the remote experience, \textit{i.e.}, using dialogue to curate exploratory experiences and providing guidance to overcome the interaction challenges faced by older adults. Prior work has studied speech interfaces for telepresence robots used by people with disabilities and observed multiple levels of abstraction of verbal commands that participants used to interact with the robot \cite{tsui_accessible_2015, tsui_designing_2013}. Echoing the findings from the prior work, we found that participants desired a combination of both low-level control and high-level instructions to control the robot. Additionally, our findings highlighted the role of the robot's dialogue beyond its use for robot control, \textit{i.e.}, supporting learning in the remote environment and providing personalized experiences.} 

\rr{First, the robot dialogue and narratives provided contextualized learning experiences for the user. We observed that participants asked a variety of questions about the remote environments, including the objects in view, local peoples' lives, the location's history, and background information. Notably, the exploratory and learning experiences are supported through a combination of the robot's narratives along with its its embodiment and interaction with the local environment. For instance, after hearing from the robot about the buoys used to collect lake and atmospheric information, one of our participants (S2.P9) requested that the robot look for the buoys during the lake visit and wanted to see them in the water. Future work should further explore how to design conversational telepresence robots to support experiential learning experiences \cite{kolb2014experiential} that are specific to the remote environment and bridge knowledge with the activities in the remote experiences.} 

\rr{Second, the robot dialogue allows the robot to ask personal questions to the user and collect their interests before or during the exploratory experience. One of our participants (S1.P3) wanted the robot to collect their preferences from her daughter who knows her well. Therefore, we suggest that the robot could involve other stakeholders such as family members and caregivers to improve the personalization of the remote experience.} 

\rr{Furthermore, having dialogue with the robot can make up for the loss of rich sensory experiences compared with the in-person experience. Technical limitations such as the camera's field of view, audio quality, and internet latency can affect the quality of conveying sensory experiences \cite{desai2011essential, heshmat_geocaching_2018}. In addition to dialogue, future design can consider enriching the narratives and knowledge of the robot to enhance the user's experiences and fill the gap of the lost sensory experiences.}


\subsubsection{Design Implications for Reminiscent Experience}
\rr{Robot dialogue also provided opportunities for participants to disclose their personal past related to the remote experience and foster reminiscence. Reminiscence has been studied as a therapeutic approach to promote older adults' mental well-being through various activities such as writing, storytelling, artistic expressions, \textit{etc.} \cite{westerhof2010reminiscence, baker2021school}.} 

\rr{Our findings reveal the rich reminiscent content that our participants recalled, including the past trips with family, college life from decades ago, and fun memories with friends. In addition to being triggered by locations of personal significance, our findings also revealed that reminiscence can be triggered by mundane objects in the remote environment. For example, our participants recalled past trips near bodies of water after seeing the lake (S1.P8, S2.P1) and recalled corsages from formal school dances when seeing the gardenia in the botanical garden (S1.P3).} 

\rr{Reminiscence often is accompanied with self-expression and disclosure about one's personal past \cite{westerhof2010reminiscence}, pointing to the need for the robot's dialogue to facilitate the reminiscence process. If visiting familiar places, the robot could ask questions about the scene in the remote environment and its association with the user. If the user begins to reminiscence, the robot could express interest in knowing more and ask follow up questions. Nevertheless, some of our participants expressed negative feedback about the appropriateness of the robot asking personal questions because either the remote location was unrelated to the participant's past experience or the participant did not want to talk about themselves with strangers. Meaningful dialogue requires trust and rapport building between the robot and the older adult \cite{ostrowski2021long}. Future design need to consider the user's privacy concerns and willingness to self-disclose before initiating conversation about personal topics. The robot need to closely monitor the the conversational topic and progress, stop in time if the user loses interests, and avoid triggering negative memories.}

\subsubsection{Design Implication for Social Participation} 




\rr{Our findings revealed the needs for the robot to take a proactive role in mediating the conversation between the local user and the remote bystander as we observed multiple communication challenges between the user and the remote bystander in study two. Often our participants could not hear the bystander well due to environment noises, especially in the outdoor space. This finding differs from prior work that suggests the telepreesence robot should take an invisible role in mediating communication\cite{tsui_accessible_2015, takayama2011toward}. We believe that that the outdoor environment in our study resulted in louder ambient noise which competed with bystander speech, compared to the quiet lab environment used in previous work \cite{tsui_accessible_2015}. The functional needs of older adults are also different, as several of our participants reported having hearing impairments. As a result, we observed challenges in turn-taking, \textit{i.e.,} our participants did not know what to say or repeatedly asked questions before the remote bystander was able to provide a response. This observation points to the need for the robot to actively facilitate communication between the user and the bystander to ease the communication challenges, such as reminding the user that the bystander is going to speak and conveying the message for the user if they can not hear clearly. }

\rr{Moreover, the background sound filter needs to consider the current state of the user's interaction, \textit{i.e.}, whether the user is talking to people or exploring in the environment. Filtering out the background sound can be beneficial during social interactions, where as the background sound can be an important and necessary component of the experience if the user is enjoying the scenery, especially in an outdoor environment.} 

\rr{Another interesting finding in our work was the violation of the social boundaries in the social engagement where the participants asked personal questions and were even rude to the bystander in the study. This observation points to the need to protect the safety of both the user and the bystander in the remote interaction, especially when they do not know each other. The robot needs to ensure both the user's and bystanders' willingness to participate in the conversation. The robot can also present a list of social norms before the interaction and provide both sides with the option to end the conversation if social boundaries are violated.}

\subsection{Limitations \& Future Work}

Our work suffers from two key limitations that might hinder the applicability of our findings and design ideas. First, our study population came from two senior living facilities located in the United States, which might not be representative of the global population of homebound older adults, as intergenerational living practices, family connections, and health and mobility services differ across geographic areas and countries. For example, homebound older adults who live in a family setting and in the neighborhood in which they spent their lives might not have the same desire for reminiscence with past experiences related to family or locations. Relatedly, our study population was primarily female. Although seven out of 10 homebound older adults are women \cite{ankuda2021association}, indicating that our sample is representative of the homebound older adult population in the United States, data from more male participants may provide insight into their needs and expectations and how technology might serve them differently. Future work must extend the geographic and gender representation of the populations we study to establish a stronger empirical foundation for an understanding of the needs of this population and their technology perceptions. 

Second, our study took a technology probe approach with a robot prototype controlled through Wizard of Oz. Technical limitations of the robot prototype such as the stability of the camera, Internet latency, and speed of the robot can negatively impact the participants' experience and cause interaction challenges, particularly given barriers to using technology homebound older adults already have. Future research should further improve the functionality of such system and address the accessibility needs of interacting with novel systems for homebound older adults.

\section{Conclusion}
In this paper, we conducted a needfinding study and a technology probe study to investigate the use of telepresence robots for homebound older adults to interact and experience the external world. From the needfinding study, we found that older adults desired reminiscent experience, exploratory experience and social participation through the telepresence robot. Then we generated design insights based on the findings and prototyped a conversational telepresence robot that can be controlled through Wizard of Oz to provide guidance and narrations about the environment, have social chats with the user, and facilitate interactions between the user and the remote bystander. Using the robot prototype as technology probe, we conducted the second study where participants remotely visited a lakefront and a botanical garden through our robot prototype. The second study revealed our participants' interaction patterns in each desired experience from the first study. Overall, this work explores the novel design space of conversational telepresence robots, specifically illustrating the potential for robot dialogue to foster more meaningful interactions while engaging in the remote experience for homebound older adults.

\section{Acknowledgments}
This work was supported in part by the McPherson Eye Research Institute and the Google Award for Inclusion Program. The robot and older adult figures in Figure 3 were designed by Freepik. We thank our participants for their time in participating in our study sessions. Thank Weinberg Terrace for helping us recruit older adults and providing the study site. Thank Phipps Conservatory and Botanical Gardens for providing the site for us to run the robot. Thank you Mrigya Kumar for helping with the project and running user studies. Special thanks to Corinne Gibson, Pearl Averbach, and Marcellina Hoskowicz for supporting us in our community partnerships.

\balance
\bibliographystyle{ACM-Reference-Format}
\bibliography{sample-base}
    
\end{document}